\documentstyle[preprint,aps]{revtex}
\input epsf.tex
\def\DESepsf(#1 width #2){\epsfxsize=#2 \epsfbox{#1}}
\begin{document}
\preprint{\vbox{\hbox{OITS-571}}}
\draft
\title{ The Nonresonant Cabibbo Suppressed Decay $B^\pm\rightarrow
\pi^+\pi^-\pi^\pm$\\
 and Signal for CP Violation}
\author{N.G. Deshpande, G. Eilam\footnote{On leave from Physics Department,
Technion- Israel Institute of Technology, 32000, Haifa, Israel.}, Xiao-Gang He,
and Josip Trampetic\footnote{On leave from Department of Theoretical Physics,
R. Bo\v{s}kovi\'{c} Institute, Zagreb 41001, Croatia}}
\address{Institute of Theoretical Science\\
University of Oregon\\
Eugene, OR 97403-5203, USA}

\date{March, 1995}
\maketitle
\begin{abstract}
We consider various contributions to the nonresonant decay $B^\pm\rightarrow
\pi^+\pi^-\pi^\pm$, both of the long-distance and short-distance types with the
former providing
for most of the branching ratio, predicted to be $BR(B^\pm\rightarrow
\pi^+\pi^-\pi^\pm) = (1.5 - 8.4 )\times 10^{-5}$. We also discuss an
application to CP violation resulting from the interference of that nonresonant
background
(with $m(\pi^+\pi^-)\approx 3.4$ GeV) and $B^\pm\rightarrow \chi_{c0} \pi^\pm$
followed by $\chi_{c0}\rightarrow \pi^+\pi^-$. The resulting value of the
partial rate asymmetry is $(0.40\sim 0.48)\mbox{sin}\gamma$, where $\gamma =
\mbox{arg}(V_{ub}^*)$.

\end{abstract}
\newpage
Two body and quasi two body non-leptonic decays of heavy mesons have been
extensively studied\cite{1}. Multibody non-leptonic decays are more difficult
to estimate, and one usually resorts to statistical or phase space
models\cite{2}. In this letter we will not discuss, for reasons that will
become clear, heavy meson decays through a chain of real resonances\cite{3},
i.e. we consider only the nonresonant background, and confine ourselves to
$B^\pm\rightarrow \pi^+\pi^-\pi^\pm$ though similar results are expected for
$B\rightarrow K K \pi$ and other modes. Our motivation is two-fold:
\begin{enumerate}
\item $B^\pm\rightarrow \pi^+\pi^-\pi^\pm$ is expected to be larger than
$B\rightarrow
\pi\pi$, which though not separated yet experimentally from $B\rightarrow
K\pi$, is estimated to have a branching ratio of the order $10^{-5}$\cite{4}.
It is therefore challenging to find a viable dynamical description of
$B\rightarrow \pi\pi\pi$.
\item Recently\cite{5},it has been suggested that large CP asymmetries should
occur in $B^\pm \rightarrow h \pi^\pm$ where the hadronic state $h =
\pi^+\pi^-$ has energy corresponding to the resonance $\chi_{c0}(3.4)$.
\end{enumerate}

The absorptive phase necessary to observe CP violation in partial rate
asymmetries, is provided by the $\chi_{c0}$ width (subtracting the small
partial width of $\chi_{c0}$ to
$\pi^+\pi^-$). The CP odd phase $\gamma$ results from the interference of the
two quark processes responsible for the background decay
$B\rightarrow\pi\pi\pi$ and $B\rightarrow \chi_{c0}\pi$, which are
$b\rightarrow u \bar u d$ and $b\rightarrow c \bar c d$, respectively. The
partial rate asymmetry obtained in Ref. \cite{5} suffers from a large
uncertainty due mostly to the unknown background and especially its angular
dependence. Note that only $h = \pi^+\pi^-$ with spin-parity $0^+$ leads to
interference with the resonant amplitude. Therefore, knowledge of the angular
dependence is crucial, and this will come out directly once one has a reliable
model for the background process $B\rightarrow \pi\pi\pi$. The interference
between the resonance and the background amplitudes will then automatically
project out the $0^+$ component of $h = \pi^+\pi^-$. Thus $\pi^+\pi^-$ arising
from resonances like $\rho$ do not interfere and need not be considered.

In this letter we will consider three contributions to $B\rightarrow \pi\pi\pi$
and identify the leading one. As demonstrated below, the branching ratio for
the background process will suffer from a large uncertainty, but the CP
violating partial rate asymmetry will be affected only mildly by this
uncertainty.

Let us now consider the three possible contributions to the nonresonant
background $B\rightarrow \pi\pi\pi$, as depicted in fig.1a-c. We choose our
momenta as follows: $B^-(p_B) \rightarrow \pi^-(p_1)\pi^+(p_2)\pi^-(p_3)$ and
always symmetrize by $p_1\leftrightarrow p_3$. Furthermore we define $s =
(p_B-p_1)^2 = (p_2+p_3)^2$ and $t = (p_B-p_3)^2 = (p_1+p_2)^2$.

Diagram 1a is
the short-distance contribution to $B\rightarrow \pi\pi\pi$, for which the
effective weak Hamiltonian is
\begin{eqnarray}
H_{eff} = {G_F\over \sqrt{2}}V_{ub}V_{ud}^* (C_1O_1 + C_2O_2)\;,
\end{eqnarray}
where $C_1 \approx -0.313$, $C_2 \approx 1.15$, and
\begin{eqnarray}
O_1 = \bar d \gamma_\mu(1-\gamma_5) b \bar u\gamma^\mu(1-\gamma_5) u\;,\;\;
O_2 = \bar u\gamma_\mu(1-\gamma_5) b \bar d \gamma^\mu (1-\gamma_5) u\;.
\end{eqnarray}
Within the factorization approximation, we have the following amplitude
\begin{eqnarray}
M_a &=& {G_F\over
\sqrt{2}}V_{ub}V_{ud}^*<\pi^+\pi^-\pi^-|C_1O_1+C_2O_2|B^->\nonumber\\
&=& {G_F\over \sqrt{2}}V_{ub}V_{ud}^*(C_1+{C_2\over N_c}) <\pi^-(p_1)|\bar d
\gamma_\mu b|B^-(p_B)>
<\pi^+(p_2)\pi^-(p_3)|\bar u\gamma_\mu u|0>\nonumber\\
& +& (p_1 \leftrightarrow p_3)
\;.
\end{eqnarray}
The matrix elements in Eq.(3), neglecting $m_\pi^2$ are
\begin{eqnarray}
<\pi^-| \bar d \gamma_\mu b|B^-> &=&
(p_B+p_1)_\mu F^{B\pi}_1(s) + {m_B^2\over s} (p_B - p_1)_\mu (F^{B\pi}_0(s) -
F^{B\pi}_1(s))\;,\nonumber\\
<\pi^+(p_2)|\bar u &\gamma_\mu& u | \pi^+(-p_3)> = (p_2 - p_3)_\mu
F_1^{\pi\pi}(s)\;.
\end{eqnarray}
Substituting in Eq.(3) and performing the scalar products lead to
\begin{eqnarray}
M_a = {G_F\over \sqrt{2}} |V_{ub}V_{ud}^*|e^{-i\gamma} a_2[
F^{B\pi}_1(s)F^{\pi\pi}_1(s)(2t+s-m_B^2)
+ F^{B\pi}_1(t)F^{\pi\pi}_1(t)(2s+t-m_B^2)]\;.
\end{eqnarray}
We have defined $a_2 = C_1+C_2/N_c$, but will take the phenomenological value
$a_2 \approx 0.24$\cite{9}, and $\gamma = \mbox{arg}(V_{ub}^*)$.
For the form factors above we use pole model forms\cite{6}
\begin{eqnarray}
F_{1,0}^{B\pi}(q^2) = {F^{B\pi}_{1,0}(0)\over 1- q^2/m_{1,0}^2}\;,\;\;
F^{\pi\pi}_1(q^2) = {1\over 1- q^2/m_{\pi\pi}^2 + i\Gamma_\sigma
/m_{\pi\pi}}\;,
\end{eqnarray}
where $F_1^{B\pi}(0) = F^{B\pi}_0(0) = 0.333$\cite{7}, or $0.53\pm0.12$\cite{8}
and
$m_1 = 5.32$ GeV, $m_0 = 5.78$ GeV, $m_{\pi\pi} \approx m_\sigma = 0.7$ GeV and
$\Gamma_\sigma = 0.2$ GeV.

Substituting the appropriate numerical values, integrating over phase space and
using \cite{10} $\tau_B = 1.54\times 10^{-12}$ s, we find that the contribution
of diagram 1a to the branching ratio is
\begin{eqnarray}
BR_a = {\Gamma_a\over \Gamma_B} = 0.9\times 10^{-6} \left ({F^{B\pi}_1(0)\over
0.333}\right )^2
\end{eqnarray}
which ranges between $0.9\times 10^{-6}$ and $2.3\times 10^{-6}$.

Diagram 1b which is obviously of the long-distance type is harder to calculate
than diagram 1a. It is nevertheless small as the intermediate pion is highly
off-shell. The weak transition $B\rightarrow \pi$ is easy to evaluate, and
leads to
\begin{eqnarray}
T(B\rightarrow \pi) &=& <\pi^-|{G_F\over
\sqrt{2}}V_{ub}V_{ud}^*(C_1O_1+C_2O_2)|B^->\nonumber\\
&=& {G_F\over \sqrt{2}} V_{ub}V_{ud}^* a_1 f_Bf_\pi m_B^2\;,
\end{eqnarray}
where $a_1 = C_1/N_c + C_2 \approx 1.1$, $f_B = 0.2$ GeV and $f_\pi = 0.13$
GeV. Then, again neglecting $m_\pi$, we find
\begin{eqnarray}
M_b = {G_F\over \sqrt{2}} V_{ub}V_{ud}^* a_1 f_Bf_\pi A(\pi\pi\pi\pi)\;.
\end{eqnarray}
$A(\pi\pi\pi\pi)$ is not known for one highly off-shell pion and three on-shell
ones. If we assume only S-wave, and use the unitarity limit, $A(\pi\pi\pi\pi)
\sim O(1)$, the branching ratio contribution of $M_b$ is
\begin{eqnarray}
BR_b = {\Gamma_b\over \Gamma_B} < 10^{-8}\;.
\end{eqnarray}
Of course it is unrealistic to assume only S-wave contribution to $M_b$, and
waves with angular momenta up to $k a$ contribute, where $k$ is the momentum in
the center of mass and $a$ is a typical size. It is difficult to make our
estimates more quantitative since one of the pions is highly off-shell. However
we can not to envision this contribution to be large, and we shall neglect
it.

Turning to diagram 1c, which is also of a long-distance type, we will show that
it is the dominant diagram and its branching ratio is equal or larger than
$BR(B\rightarrow \pi\pi)$ which should clearly be the case, since even in the
charmed meson system\cite{10} $\Gamma(D\rightarrow \pi\pi\pi)\geq
\Gamma(D\rightarrow\pi\pi)$. The calculation of the amplitude $M_c$ involves
the application of both Heavy Quark Effective Theory (HQET) and Chiral
Perturbation Theory (CHPT).
For a review of both see Ref.\cite{11}. First we write
\begin{eqnarray}
M_c = A^\mu_{BB^*\pi} {-g_{\mu\nu} + p_{B^*\mu}p_{B^*\nu}/m^2_{B^*}\over
p^2_{B^*} - m_{B^*}^2} A^\nu_{B^*\pi\pi} + (p_1\leftrightarrow p_3)\;.
\end{eqnarray}
Note that the $B^*$ is off-shell and since we are interested in the nonresonant
part of $B\rightarrow\pi\pi\pi$, no on-shell intermediate resonances are
introduced. Our main aim now is to calculate the strong and weak vertices
$A^\mu_{BB^*\pi}$ and $A^\nu_{B^*\pi\pi}$, respectively, using the methods of
HQET and for the strong vertex combining them with CHPT\cite{12}.

Let us start by calculating $A^\mu_{BB^*\pi}$. The Heavy-Chiral Lagrangian
density\cite{11,12} relevant to us is
\begin{eqnarray}
\cal{L}\mbox{$_{int} = ig \sqrt{m_Bm_{B^*}} < H_b \gamma_\mu\gamma_5 A^\mu_{ba}
\bar H_a>$}\;,
\end{eqnarray}
where $<>$ stands for trace. The field $H_a$ describes the heavy-quark
light-quark
$(Q\bar q_a)$ system and
\begin{eqnarray}
H_a &=& {1+\not v \over 2} (P^*_{a\mu} \gamma^\mu - P_a\gamma_5)\;,\;\;
\bar H_a = \gamma_0 H^\dagger \gamma_0\;,\nonumber\\
A^\mu_{ba} &=& {1\over 2}(\xi^\dagger \partial^\mu \xi - \xi \partial^\mu
\xi^\dagger)_{ba}\;,
\end{eqnarray}
where $P_a = (B^-, \bar B^0_d, \bar B^0_s)$ and similarly for $P^*_{a\mu}$ in
terms of the vector meson states, $v$ is the heavy meson velocity, and $\xi =
exp(iM/f_\pi)$ with $M$ given by
\begin{eqnarray}
M = \left ( \begin{array}{ccc}
{\pi^0\over \sqrt{2}}+{\eta\over \sqrt{6}}& \pi^+&K^+\\
\pi^-&-{\pi^0\over \sqrt{2}}+{\eta\over \sqrt{6}}& K^0\\
K^-&\bar K^0& - {2\eta\over \sqrt{6}}
\end{array}
\right )\;.
\end{eqnarray}
We obtain
\begin{eqnarray}
A_{BB^*\pi}^\mu\epsilon_\mu = -{2g\over f_\pi} \sqrt{m_Bm_{B^*}}B^-B^*_\nu
\partial ^\nu
\pi^+\;.
\end{eqnarray}
Using the flavor symmetry of HQET the coupling constant $g$ is determined to be
0.6 from
$D^*\rightarrow D\pi$ data\cite{8,12}. The main uncertainty in the application
of Eq.(15) to our case is that in diagram 1c the $B^*$ is off-shell. We
therefore define $\mu$ as a measure of the off-shellness of the $B^*$ and
consider two cases: 1. $\mu = \sqrt{m_B m_{B^*}}$ in Eq.(15). 2. $\mu =
\sqrt{m_B\sqrt{p_{B^*}^2}}$, where $p_{B^*}$ is the momentum of the $B^*$.

To calculate $A_{B^*\pi\pi}$ in Eq.(11), we employ the
spin independence of HQET and write
\begin{eqnarray}
A_{B^*\pi\pi}^\nu\epsilon_\nu &=& {G_F\over
\sqrt{2}}V_{ub}V_{ud}^*<\pi^+\pi^-|C_1O_1+C_2O_2|B^*>\nonumber\\
&=&{G_F\over \sqrt{2}}V_{ub}V_{ud}^* a_1<\pi^+|\bar u \gamma_\mu (1-\gamma_5)
b|B^*> <\pi^-|\bar d \gamma^\mu(1-\gamma)u|0>\;.
\end{eqnarray}
The form factors $T_{1-4}$ are defined as follows
\begin{eqnarray}
<\pi^+|\bar u\gamma_\mu b |B^*> &=& 2 T_1 i
\epsilon_{\mu\nu\lambda\sigma}\epsilon^\nu p^\lambda_{B^*}
p^\sigma_2\;,\nonumber\\
<\pi^+|\bar u\gamma_\mu \gamma_5 b|B^*>&=& 2T_2 m_{B^*}^2 \epsilon_\mu +
2T_3 (\epsilon\cdot q)(p_{B^*}+p_2)_\mu +2T_4(\epsilon\cdot q)
(p_{B^*}-p_2)_\mu\;,
\end{eqnarray}
where $q = p_{B^*}-p_2$. Relations between $T_i's$ and $f_\pm$ defined through
\begin{eqnarray}
<\pi^+&|&\bar u\gamma_\mu b|\bar B^0> = f_+ (p_B +p_\pi)_\mu
+f_-(p_B-p_\pi)_\mu\;,
\end{eqnarray}
are
\begin{eqnarray}
T_1 &=& -i{f_+-f_-\over 2m_B}\;,\;\; T_2 = {1\over 2 m_B^2}((f_++f_-)m_B +
(f_+-f_-){p_B\cdot p_\pi\over m_B})\;,\nonumber\\
T_3 &=& {f_+-f_-\over 4m_B}\;,\;\; T_4 = T_3\;.
\end{eqnarray}
Substituting the above relations in Eq.(16), we have
\begin{eqnarray}
A_{B^*\pi\pi}^\nu\epsilon_\nu = {G_F\over \sqrt{2}}V_{ub}V_{ud}^*a_1 f_\pi
(\epsilon\cdot p_3) (
{3f_+\over 2}m_B + {f_-\over 2}m_B +(f_+-f_-){p_2\cdot p_3\over m_B})\;.
\end{eqnarray}
The amplitude for diagram 1c, obtained from Eq.(11), (15) and (20) expressed
in terms of $F^{B\pi}_{1,0}$
\begin{eqnarray}
M_c &=& -{G_F\over \sqrt{2}}V_{ub}V_{ud}^*(2g a_1) F^{B\pi}_1(m_\pi^2)
{\mu\over s-m_{B^*}^2} [{3\over 2}m_B + {s\over 2m_B^2}\nonumber\\
&+& {m_B\over 2} {m_B^2-s\over m^2_\pi}(1 - {F^{B\pi}_0(m_\pi^2)\over
F^{B\pi}_1(m^2_\pi)})][-{s^2\over 4 m_B^2} + ({1\over 2}+{m_B^2\over 4
m_{B^*}^2})s + {t-m_B^2\over 2}]
+ (s\leftrightarrow t)\;.
\end{eqnarray}
The branching ratio implied by diagram 1c is
\begin{eqnarray}
BR_c = {\Gamma_c\over \Gamma_B} = \left \{ \begin{array}{ll}
3.3\times  10^{-5}({F^{B\pi}_1(0)\over 0.333})^2\;,& \mbox{case
1}\;,\nonumber\\
1.5\times 10^{-5}({F^{B\pi}_1(0)\over 0.333})^2\;,& \mbox{case 2}\;.
\end{array}
\right .
\end{eqnarray}
thus obtaining $BR_c = (1.5\sim 8.4)\times 10^{-5}$. The spread is caused by
the two different prescriptions for taking into account the off-shellness of
the $B^*$, and by the fact that $0.333 \leq F^{B\pi}_1(0) \leq 0.53$.
Since $BR_c$ is the largest branching ratio as compared to $BR_a$ and $BR_b$,
and is not smaller than the branching ratio for $B\rightarrow \pi\pi$, we take
$BR_c$ as a good estimate for the branching ratio of the nonresonant decay
$B\rightarrow \pi\pi\pi$, and obviously
$M(B\rightarrow \pi\pi\pi) = M_c$.

It is not surprising that three-body decays are dominated by a long-distance
contribution in contrast to the two-body decays which are dominated by
factorization and a short-distance amplitude. The mechanism of producing
additional pions
must necessarily involve the strong interaction.

Turning now to the CP violating asymmetry, we interfere $M_c$ with the
resonance amplitude $M_{res}$ for $B^\pm\rightarrow \chi_{c0}\pi^\pm
\rightarrow \pi^+\pi^-\pi^-$ from diagram 1d, where
\begin{eqnarray}
M_{res} = A(B^\pm \rightarrow \chi_{c0}\pi^\pm) {1\over
s-m_\chi^2 +i\Gamma_\chi m_\chi} A(\chi_{c0} \rightarrow \pi^+\pi^-) +
(s\leftrightarrow t)\;.
\end{eqnarray}
Following Ref. \cite{5} we integrate the decay rate in the phase space from
$s_{min} =
(m_\chi - 2\Gamma_\chi)^2$ to
 $s_{max} = (m_\chi + 2\Gamma_\chi)^2$ where $m_\chi$ and $\Gamma_\chi$ are the
mass
and width, respectively of $\chi_{c0}$. We define the partial width
$\Gamma_p \sim \int ds dt |M_c + M_{res}|^2$, where $0\leq t \leq m_B^2-s$ and
the $s$ integral has the above limits.
Therefore the absolute value of the asymmetry
\begin{eqnarray}
|A| = \left | {\Gamma_p - \bar \Gamma_p\over \Gamma_p +\bar \Gamma_p} \right |
= (0.40\sim 0.48)\mbox{sin}\gamma\;.
\end{eqnarray}
Since, unlike the case for Ref.\cite{5}, where the amplitude for the
nonresonant background is unknown as a function of both s and t (and therefore
its angular dependence is unknown), here the model used dictates the angular
dependence which gives more confidence in the asymmetry obtained.
It is interesting that the large uncertainty in the background $BR(B\rightarrow
\pi\pi\pi)$ does not translate into a large spread in the values for $|A|$
since it affects both numerator and denominator in $|A|$. From the very large
direct CP violation asymmetry obtained for $\mbox{sin}\gamma = 1$ and using
$BR(B^\pm\rightarrow \chi_{c0}\pi^-)BR(\chi_{c0}\rightarrow \pi^+\pi^-) \approx
5 \times 10^{-7}$,
the number of events N required experimentally to detect such an asymmetry at
the $3\sigma$ level is $9\times 10^{7} \leq N \leq 13 \times 10^{7}$. One
expects future B factories to be able to reach such a number of events.

Finally let us note that other modes of the $\chi_{c0}$ are suitable for
similar considerations, in particular $\chi_{c0}\rightarrow K K$ for which we
expect more or less the same result for the asymmetry in $B\rightarrow K K
\pi$. Even larger CP violation asymmetries are expected for $B\rightarrow h
\pi$ where now $h = 2(\pi^+\pi^-)\;, \pi^+\pi^- K^+K^-$ for which
$BR(\chi_{c0}\rightarrow h)$ is at a level of a few percent. Estimates of the
nonresonant background unfortunately become more difficult. The same situation
(large asymmetry, but difficult to predict the nonresonant background) is
expected in $B\rightarrow h\pi$ where $h = \eta'
\pi\pi\;, \rho\rho$ etc., and the nonresonant amplitude interferes with
$B\rightarrow \eta_c \pi$.

\acknowledgments
This work has been supported in part by the Department of Energy Grant No.
DE-FG06-85ER40224. The work of G.E. has been supported in part by the
Binational Science Foundation Israel-US and by the VPR fund. The work of J.T.
was supported in part by the Croatian Ministry of Research under contract
1-03-199. Both G.E. and J.T would like to thank the members of the Institute of
Theoretical Science for their warm hospitality.

\begin{figure}[htb]
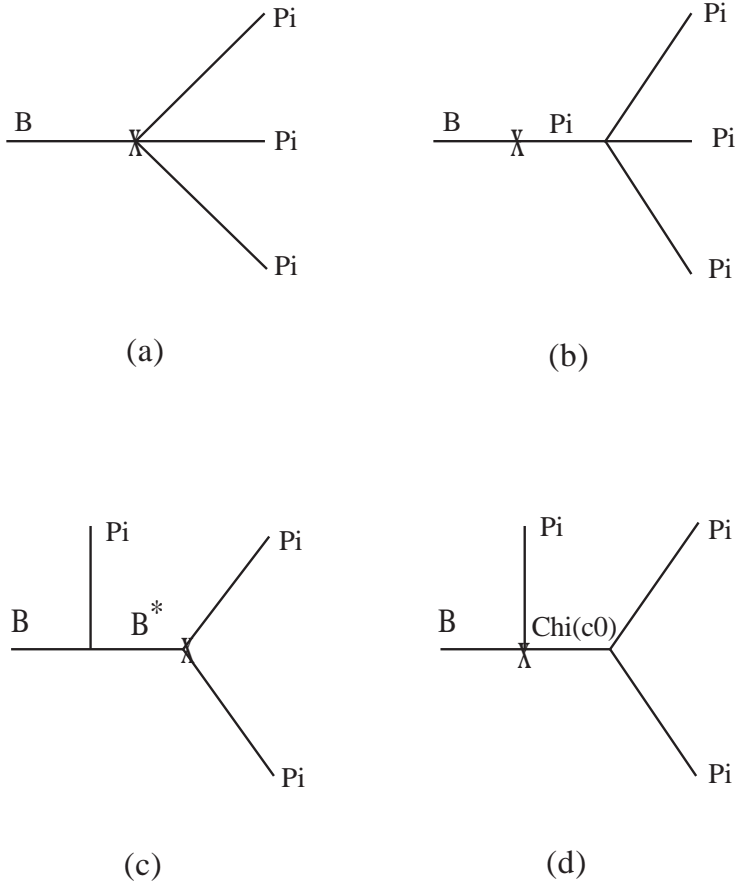

\centerline{ \DESepsf(gad.epsf width 10 cm) }
\smallskip
\caption{Diagrams contributing to $B^\pm\rightarrow \pi^+\pi^-\pi^\pm$. In
these diagrams weak vertices are indicated by X. }
\label{gad}
\end{figure}

\end{document}